\documentclass[aps,prl,twocolumn,showpacs,preprintnumbers,amsmath,amssymb,superscriptaddress]{revtex4-1}

\usepackage{graphicx}
\usepackage{dcolumn}
\usepackage{color}
\usepackage{bm}     
\usepackage{amsmath}
\usepackage[breaklinks,colorlinks,bookmarks=false,citecolor=blue,linkcolor=red,urlcolor=blue]{hyperref}

\begin{document} 

\title{The Higgs Mode of Planar Coupled Spin-Ladders\\ and its Observation in  C$_9$H$_{18}$N$_2$CuBr$_4$}

\author{T. Ying}
\affiliation{Institut f\"ur Theoretische Festk\"orperphysik, JARA-FIT and JARA-HPC, RWTH Aachen University, 52056 Aachen, Germany}
\affiliation{Department of Physics, Harbin Institute of Technology, 150001 Harbin, China}

\author{K. P. Schmidt}
\affiliation{Institut  f\"ur Theoretische Physik, FAU Erlangen-N\"urnberg, Germany}

\author{S. Wessel}
\affiliation{Institut f\"ur Theoretische Festk\"orperphysik, JARA-FIT and JARA-HPC, RWTH Aachen University, 52056 Aachen, Germany}

\date{\today}

\begin{abstract}
Polarized inelastic neutron scattering experiments recently identified the amplitude (Higgs) mode in C$_9$H$_{18}$N$_2$CuBr$_4$, a
two-dimensional near-quantum-critical  spin-1/2  two-leg ladder compound, which exhibits a weak easy-axis exchange anisotropy. Here, we theoretically examine the dynamic spin structure factor of such planar coupled spin-ladder systems using large-scale quantum Monte Carlo simulations. This allows us to provide a quantitative account of   the experimental neutron scattering data within a consistent quantum spin model.
Moreover, we  trance the details of   the continuous evolution of the amplitude mode from a two-particle bound state of coupled ladders in the classical Ising limit all the way to the quantum  spin-1/2 Heisenberg limit with fully restored SU(2) symmetry, where it gets overdamped by the two-magnon continuum in neutron scattering. 
\end{abstract}

\maketitle

A central aspect of current research in quantum magnetism is the exploration of emerging phases and quantum phase transition and the associated  collective excitations of quantum matter.
For one of the most fundamental ordering phenomena in quantum magnetism
 -- antiferromagnetism from spontaneous  SU(2) spin symmetry breaking -- the 
collective excitations can be characterized as fluctuations in the phase and the amplitude of the order parameter field. The phase oscillations correspond to low-energy magnon modes, i.e., gapless Nambu-Goldstone bosons,
which are  readily detected  in inelastic neutron scattering (INS) experiments.  However, in low-dimensional systems, for which quantum fluctuations prevail,
the  Higgs mode, associated to the amplitude fluctuations, is prone to decay into pairs of  Nambu$-$Goldstone modes~\cite{Podolsky11,Podolsky12,Pekker15}. In low-dimensional magnets, the Higgs mode  thus gets strongly masked by this coupling to the two-magnon continuum, which makes its detection formidable by magnetic probes such as  INS~\cite{Affleck92,Weidinger15}.  
However, near-quantum-critical systems were recently found to be providential for the  detection of the Higgs  mode in 2D systems, alert via its response in scalar susceptibilities as opposed to the magnetic response accessed in, e.g., INS experiments~\cite{Podolsky11,Podolsky12,Pollet12,Chen13,Gazit14,Rancon14,Lohoefer17}. 
 
 \begin{figure}[t]
\includegraphics[width=0.9\columnwidth]{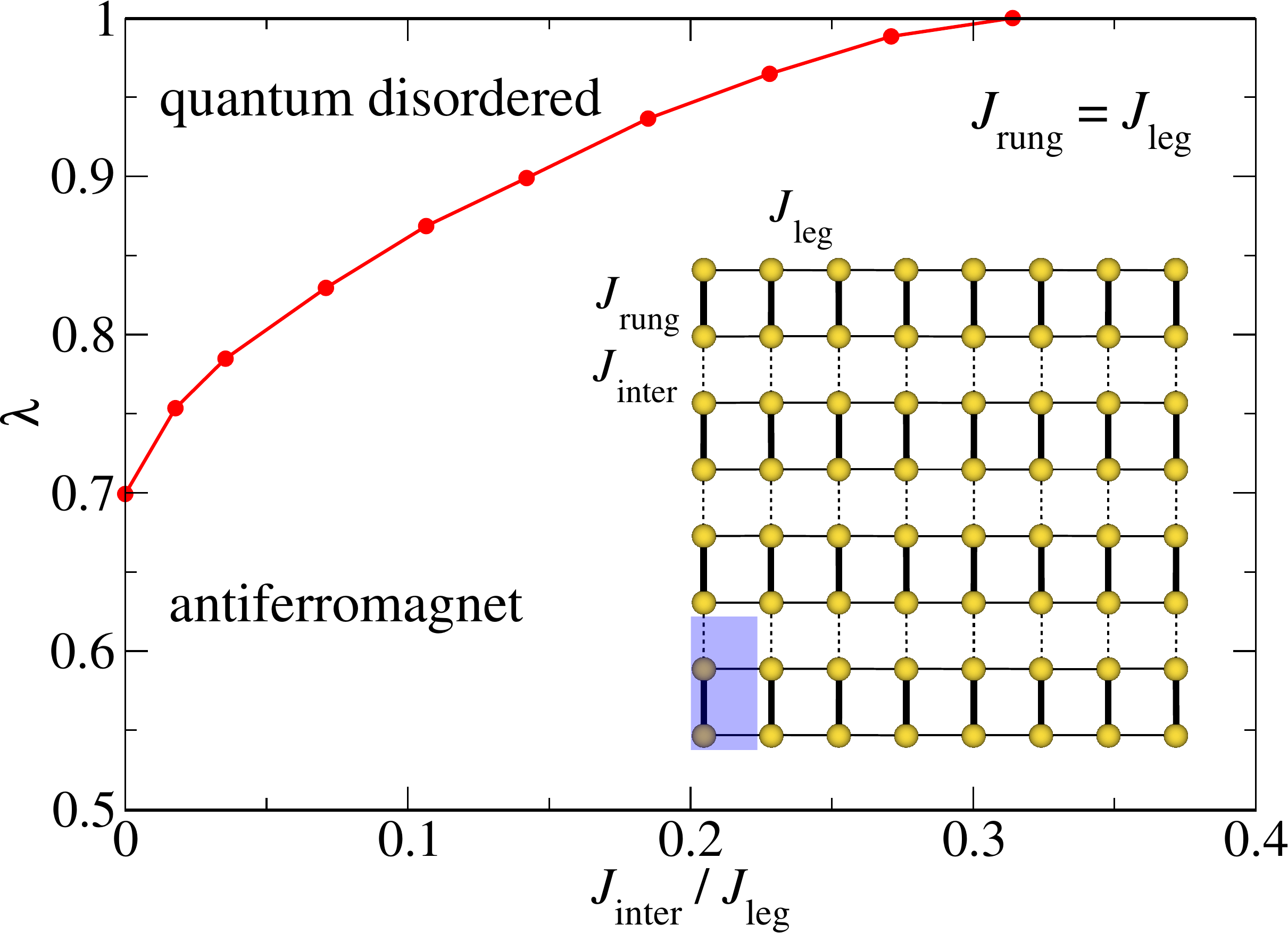}
\caption{Ground-state phase diagram of a 2D array of coupled spin-1/2 ladders with  easy-axis anisotropy $\lambda$ and $J_\mathrm{rung}=J_\mathrm{leg}$.  Inset: 2D array of coupled ladders ($L=4$ ladders, $L_r=8$ rungs per ladder), with a two-site unit cell as indicated.}
\label{fig:phasediag}
\end{figure}

A feasible route towards  the observation of the Higgs mode in near quantum-critical low-dimensional magnets 
 was explored in a recent INS study~\cite{Hong17} of the layered system of coupled  spin-ladders in  the metall-organic compound C$_9$H$_{18}$N$_2$CuBr$_4$, abbreviated as DLCB. In this compound, the spin-1/2 degrees of freedom on the Cu$^{2+}$ ions experience a weakly anisotropic, easy-axis  spin-exchange interaction~\cite{Hong14}. This anisotropy gaps out  the two-magnon scattering continuum sufficiently above the spectral  support of the lower lying Higgs mode,
 which acquires an infinite lifetime.  The Higgs mode can thus be identified by spin-polarized INS through the longitudinal, (non-spin-flip) channel, where the neutrons' polarization is vertical to the scattering plane,  separated from the magnon branch in the transverse (spin-flip) channel~\cite{Hong17}. A 2D array of coupled spin-ladders furthermore exhibits a line of quantum critical points in a parameter regime that separates the antiferromagnetic ground state from  the quantum disordered regime at weak inter-ladder coupling~\cite{Matsumoto01}. Being located near such a quantum critical point, a quantitative theory of the quantum spin dynamics in DLCB requires an approach that accounts for both the enhanced quantum critical fluctuations as well as the subtle energetics of the weakly anisotropic exchange. 

Here, we demonstrate  such a quantitative theoretical characterization of the quantum spin dynamics in  coupled spin-ladders with anisotropic exchange: 
Given the absence of geometric frustration in the exchange geometry derived for DLCB~\cite{Hong14,Hong17},  an unbiased approach for calculating the dynamic spin structure  factor (DSF) is shown to be feasible using state-of-the-art quantum Monte Carlo (QMC) methods. In addition to  modeling  the INS experiments on DLCB, we  harness  the QMC approach in order to systematically examine the evolution of the magnetic excitations from the isotropic (Heisenberg) limit with its full  SU(2) symmetry, down to the Ising-model limit for dominant easy-axis exchange. The Higgs mode, which becomes overdamped in the Heisenberg limit, then connects to a gapped two-magnon bound state in the Ising-model regime. In contrast, for weakly coupled ladders, the same mode instead condenses, and gives rise to a quantum disordered phase.

In the following, we consider as a minimal  model~\cite{Hong14} for DLCB the quantum spin-1/2 Hamiltonian of a 2D array of coupled two-leg spin-ladders, 
\begin{eqnarray}\label{eq:model}
&H=&J_\mathrm{rung} \!\sum_{i,r} \! \lambda( S^x_{i,r,1}  S^x_{i,r,2}  +   S^y_{i,r,1}  S^y_{i,r,2})  \!+ \! S^z_{i,r,1}  S^z_{i,r,2}\\
&+&\!\!\!J_\mathrm{leg} \!\sum_{i,r,l} \!\lambda (S^x_{i,r,l}  S^x_{i,r+1,l} + S^y_{i,r,l}  S^y_{i,r+1,l}) \! + \! S^z_{i,r,l}  S^z_{i,r+1,l}\nonumber\\
&+&\!\!\!J_\mathrm{inter}\! \sum_{i,r} \! \lambda (S^x_{i,r,2}  S^x_{i+1,r,1}  \!+ S^y_{i,r,2}  S^y_{i+1,r,1})  \!+ \! S^z_{i,r,2}  S^z_{i+1,r,1}\nonumber
\end{eqnarray}
where $i$ indexes the ladders,  $r$ the rungs, and $l=1,2$ the two legs of each ladder. $J_\mathrm{inter}$ denotes the nearest-neighbor interladder coupling, and $J_\mathrm{leg}$  ($J_\mathrm{rung}$)  the intra-ladder couplings along the legs (rungs), respectively (cf. the inset of Fig.~\ref{fig:phasediag}).  Furthermore, $\lambda$ is the exchange anisotropy, with
$0\leq \lambda<1$ in the easy-axis regime,  which is considered equal 
among all exchange interactions~\cite{Hong14}. The Heisenberg limit is recovered at $\lambda=1$,  while for $\lambda=0$, $H$ reduces to a classical Ising model. 
An explicit constraint on the  parameters in Eq.~(\ref{eq:model}) for DLCB  follows 
from its magnetic saturation field of  
$H_\mathrm{sat}\approx16 \:\mathrm{T}$, i.e., 
\begin{equation}\label{eq:constraint}
\frac{1+\lambda}{2}(J_\mathrm{rung}+2J_\mathrm{leg}+J_\mathrm{int})=g \mu_B H_\mathrm{sat}
\approx 1.96 \:\mathrm{meV},
\end{equation}
based on a value of $g=2.12$~\cite{Hong16}. From comparing the low-temperature  INS spectra to  magnon dispersions  obtained within a perturbative continuous unitary transformation (pCUT) approach, Ref.~\onlinecite{Hong14} reports the best-fit values $J_\mathrm{rung}=0.64(9)\:\mathrm{meV}$, $J_\mathrm{leg}=0.60(2)\:\mathrm{meV}$,  $J_\mathrm{inter}=0.19(2)\:\mathrm{meV}$, and $\lambda=0.93(2)$.
These parameters position DLCB  close to  quantum criticality, where the long-range antiferromagnetic order 
along the easy-axis direction 
vanishes: In the Heisenberg limit ($\lambda=1$) for  spatially isotropic ladders  ($J_\mathrm{rung}=J_\mathrm{leg}$), this quantum critical point is located at a critical ratio of 
$J_\mathrm{inter}/J_\mathrm{leg}=0.31407(5)$~\cite{Matsumoto01}. 
The value  of $\lambda<1$ is in accord with the constraint in Eq.~(\ref{eq:constraint}), and 
accounts for the finite excitation gaps 
$\Delta_\mathrm{TM}=0.33(3)\:\mathrm{meV}$,
and $\Delta_\mathrm{LM}=0.48(3)\:\mathrm{meV}$, 
estimated in polarized INS for the transverse magnon mode (TM) and the longitudinal Higgs mode (LM), respectively~\cite{Hong17}. 
A finite $\Delta_\mathrm{TM}$ not only renders the Nambu-Goldstone mode from the  
isotropic case massive, it also leads to a minimum excitation energy of $2\Delta_\mathrm{TM}$ for the two-magnon continuum. 
For $\Delta_\mathrm{LM}<2\Delta_\mathrm{TM}$, 
the Higgs mode is protected against decay into the two-magnon continuum, thus allowing for its identification in the  longitudinal scattering channel~\cite{Hong17}. The theoretical modeling of the INS data in this configuration was performed in 
Ref.~\onlinecite{Hong17} using bond-operator theory (BOT) in harmonic approximation~\cite{Sachdev90, Sommer01}. However, within this mean-field treatment, the 
comparison to the experimental data required a substantial renormalization of the exchange couplings in the Hamiltonian 
of Eq.~(\ref{eq:model}),   up to   factors of almost two,  compared to the values quoted above. 
This  calls for an unbiased, consistent theoretical understanding 
of the INS  results on DLCB, which applies to \textit{both} scattering channels, and  also  accounts  for  the critically enhanced quantum fluctuations.

\begin{figure}[t]
\includegraphics[width=0.9\columnwidth]{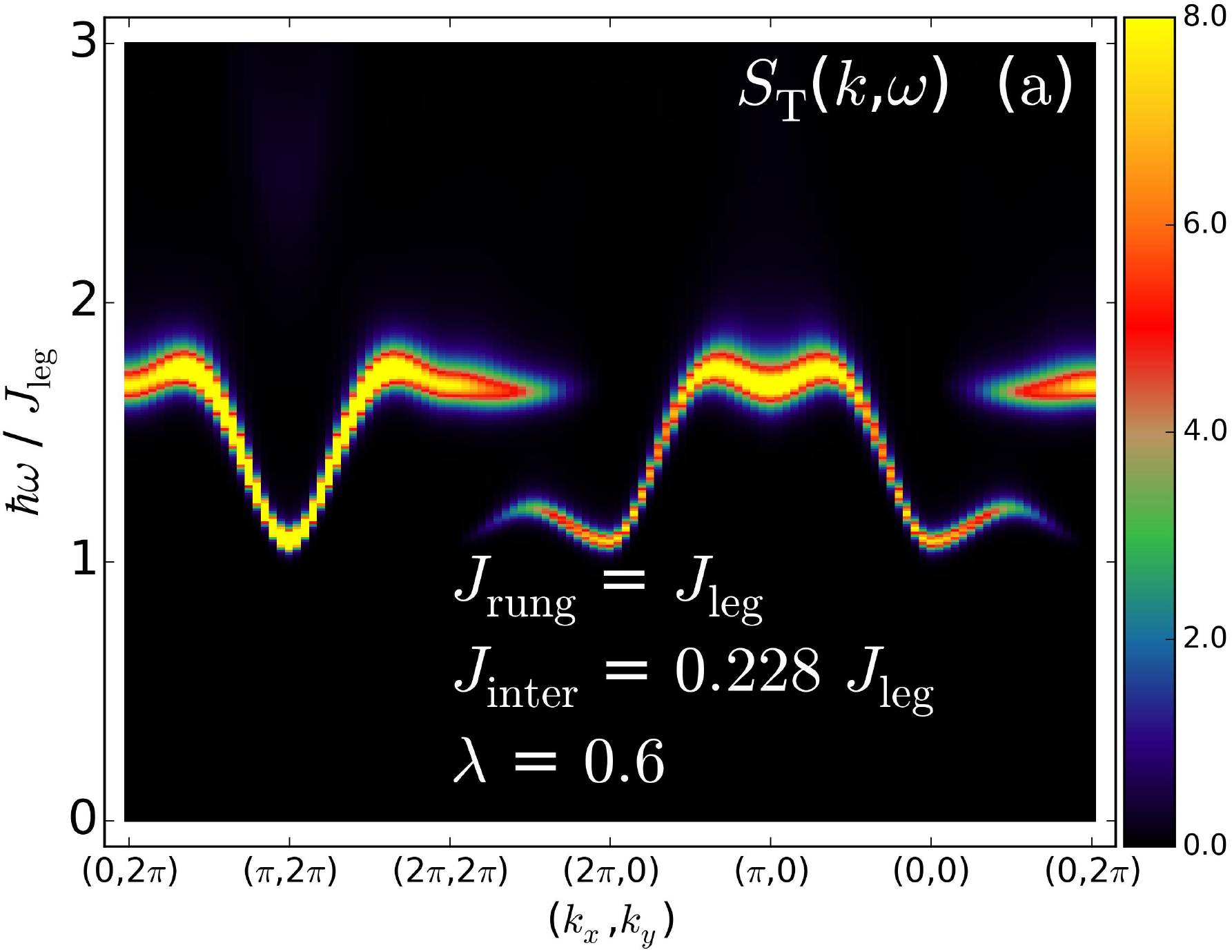}
\includegraphics[width=0.9\columnwidth]{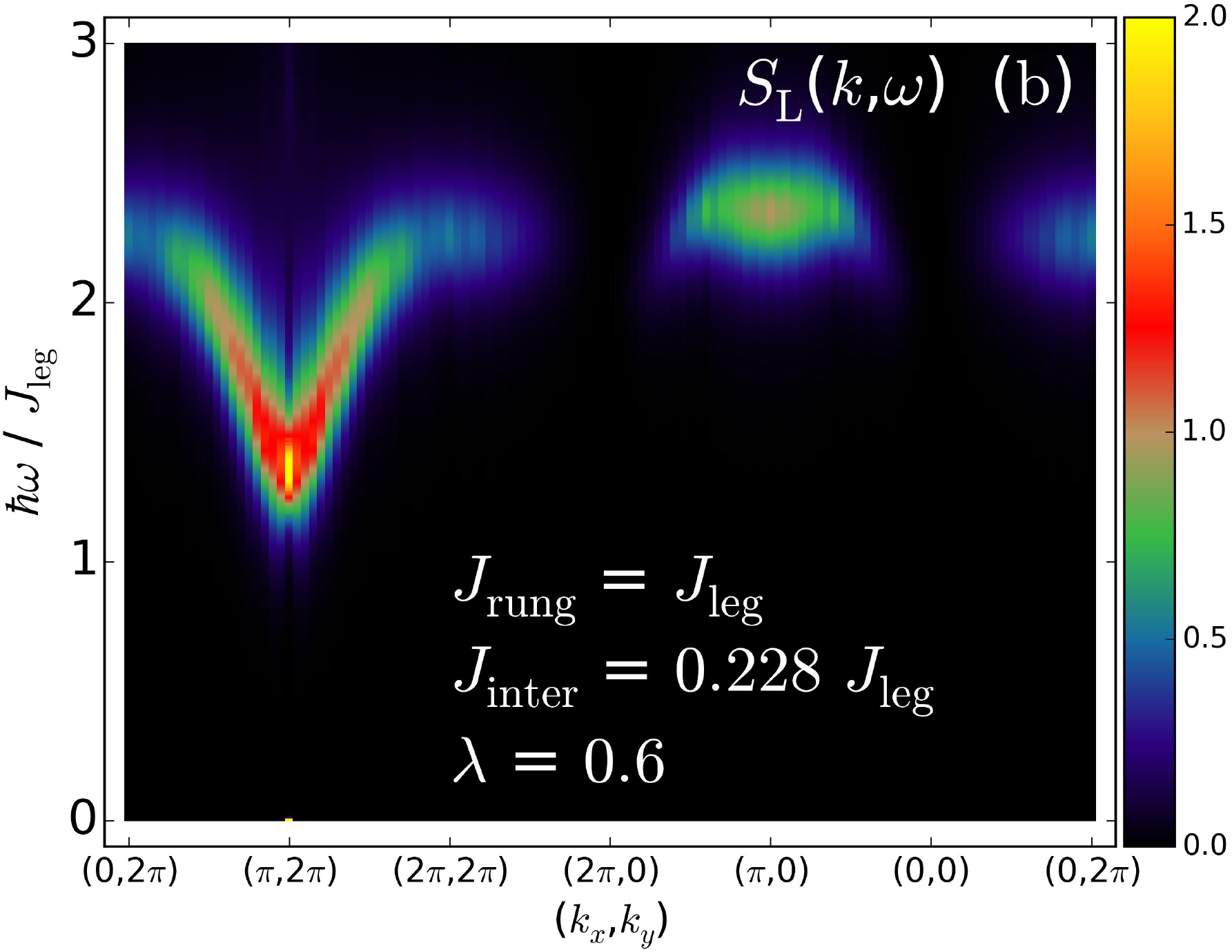}
\caption{DSF of a 2D array of coupled spin-ladders,  
(a) $S_\mathrm{T}(\bm{k},\omega)$,  and 
(b) $S_\mathrm{L}(\bm{k},\omega)$,  
for $J_\mathrm{rung}=J_\mathrm{leg}$, \mbox{$J_\mathrm{inter} =0.228 J_\mathrm{leg}$}, and $\lambda=0.6$, along the indicated path in momentum space, obtained by QMC with $L=20$ at $T=0.02 J_\mathrm{leg}$.
}
\label{fig:skw}
\end{figure}

For this purpose, we analysed the DSF of the Hamiltonian $H$ using a combination of QMC
simulations~\cite{Sandvik99,Syljuasen02,Alet05,Michel07} and a stochastic analytical continuation scheme~\cite{Beach04} in order to access the frequency-dependent spectral functions from  imaginary-time correlation functions obtained by the QMC calculations.  We thereby  obtain the DSF for both the longitudinal channel,
 $S_\mathrm{L}(\bm{k},\omega) = \int\mathrm{d}t \, \mathrm{e}^{-\mathrm{i}\hbar\omega t} 
 \langle {S}^z_{\bm{k}}(t) {S}^z_{-\bm{k}}(0)  \rangle$,
as well as for the transverse channel,
 $S_\mathrm{T}(\bm{k},\omega) = \int\mathrm{d}t \, \mathrm{e}^{-\mathrm{i}\hbar\omega t} 
 \langle {S}^+_{\bm{k}}(t) {S}^-_{-\bm{k}}(0) +  {S}^-_{\bm{k},}(t) {S}^+_{-\bm{k}}(0)\rangle$~\cite{SM}. Here,
$\bm{S}_{\bm{k}}=\frac{1}{\sqrt{N}} \sum_i \mathrm{e}^{-\mathrm{i} \bm{k}\cdot \bm{r}_{i} } \bm{S}_{i}$, and $N$ denotes the number of spins, with $N=2 \: L \: L_r$ in terms of the number of ladders ($L$) and rungs per ladder ($L_r$), with periodic boundary conditions taken in both lattice directions (the unit cell contains two spins, cf. the inset of Fig.~\ref{fig:phasediag}, and the extend of the unit cell is  set 
equal to  unity in both lattice directions). For the QMC simulations,  performed using the stochastic series expansion approach~\cite{Sandvik99,Syljuasen02,Alet05}, we scaled $L_r=2L$ and the temperature $T$ sufficiently low  to  access ground state properties of these finite systems~\cite{SM}.
 
Prior to focusing on DLCB, we  consider the simpler case of  spatially isotropic ladders ($J_\mathrm{rung}=J_\mathrm{leg}$), for which the ground-state phase diagram in terms of the ratio $J_\mathrm{inter}/J_\mathrm{leg}$ and $\lambda$, as obtained from QMC simulations,  is shown in Fig.~\ref{fig:phasediag}. In addition to a phase with antiferromagnetic order, this phase diagram exhibits an extended quantum disordered regime at weak interladder coupling near the Heisenberg limit. For $\lambda<1$, a line of quantum critical points  separates both phases, belonging to the three-dimensional (3D) Ising universality class,   in accord with a  standard finite-size scaling analysis  of the antiferromagnetic structure factor~\cite{SM}.
For $\lambda=1$, the quantum critical point at $J_\mathrm{inter}/J_\mathrm{leg}=0.31407(5)$ instead belongs to the 3D Heisenberg universality class~\cite{Matsumoto01}.
 
We now examine in detail the evolution of the DSF upon tuning $\lambda$ for $J_\mathrm{inter}/J_\mathrm{leg}=0.228$, and $0.4$, i.e., on both sides of the  critical coupling ratio for  $\lambda=1$. These two different regimes are denoted as case I and II, respectively. As an example,  
Fig.~\ref{fig:skw} displays the DSF  for $J_\mathrm{inter}/J_\mathrm{leg}=0.228$ and $\lambda=0.6$, along the indicated path in momentum space that includes the antiferromagnetic ordering vector $\bm{k}_\mathrm{AF}=(\pi,2\pi)$. The  transverse channel, $S_\mathrm{T}(\bf{k},\omega)$, is dominated by the gapped magnon excitation, with a minimum gap $\Delta_\mathrm{TM}\approx 1.1 J_\mathrm{leg}$  at $\bm{k}_\mathrm{AF}$. This sets the lower threshold for the two-magnon continuum to $2\Delta_\mathrm{TM}\approx 2.2  J_\mathrm{leg}$.  Besides the magnetic Bragg peak at $\bm{k}_\mathrm{AF}$, $S_\mathrm{L}(\bf{k},\omega)$ exhibits an additional, pronounced dispersing mode  at energies significantly below  $2\Delta_\mathrm{TM}$, and with a corresponding minimum gap of 
$\Delta_\mathrm{LM}\approx 1.3 J_\mathrm{leg}$ at $\bm{k}_\mathrm{AF}$. Its origin becomes explicit in the Ising limit: For $\lambda=0$, the ground states are perfect N\'eel configurations, and a single spin flip costs an excitation energy $\Delta_\mathrm{TM}=(J_\mathrm{rung}+2J_\mathrm{leg}+J_\mathrm{int})/2$.  A bound state of two nearest-neighbor spin flips along an intra-ladder bond (for $J_\mathrm{rung}=J_\mathrm{leg}>J_\mathrm{int}$) requires an energy 
$\Delta_\mathrm{LM}=2J_\mathrm{leg}+J_\mathrm{int}$,
which falls below the excitation energy $2\Delta_\mathrm{TM}$ for two isolated spin flips. The transverse exchange for finite values of $\lambda$ renders these modes  dispersive, thereby reducing  both excitation gaps. 

 \begin{figure}[t]
\includegraphics[width=0.492\columnwidth]{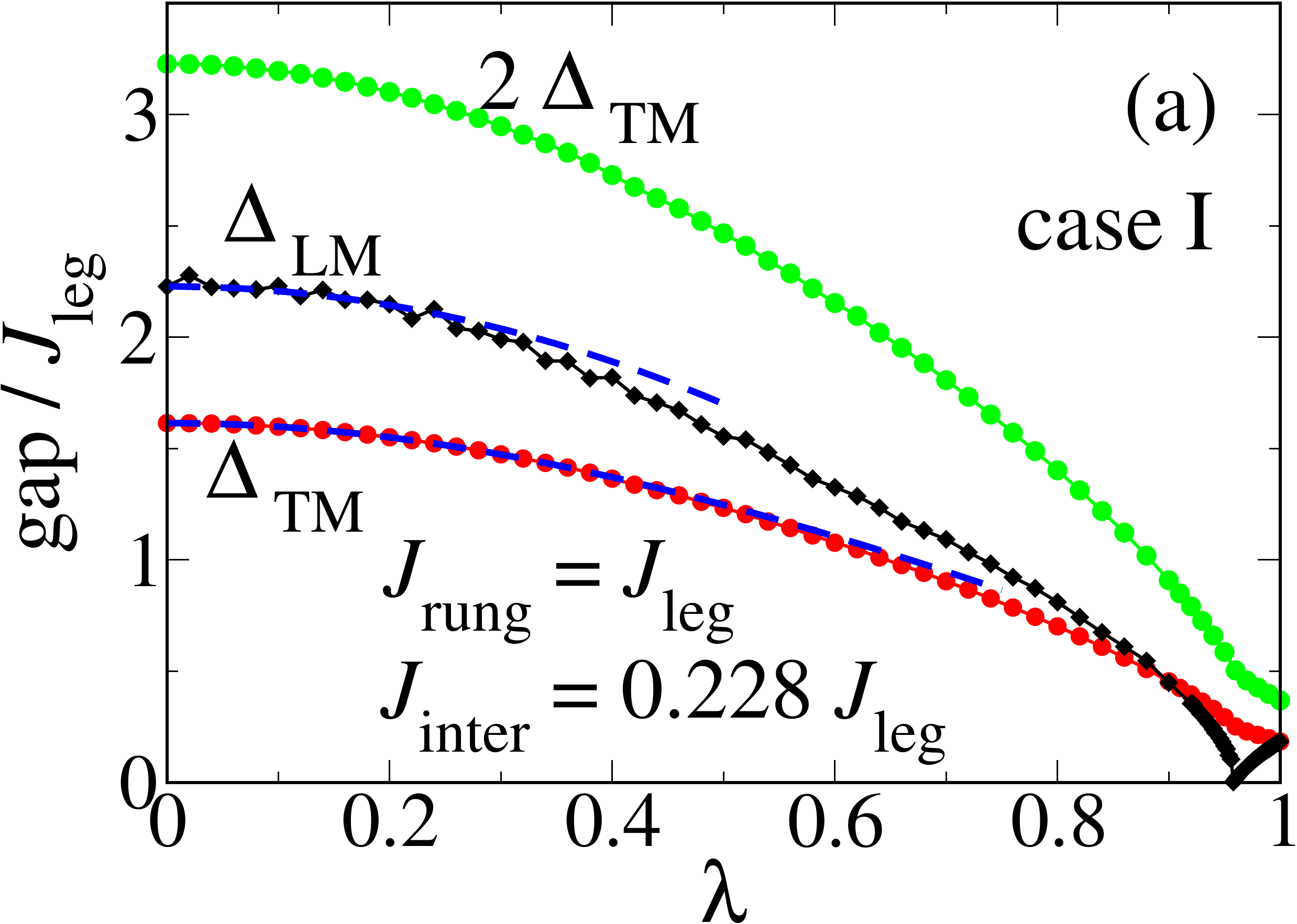}
\includegraphics[width=0.492\columnwidth]{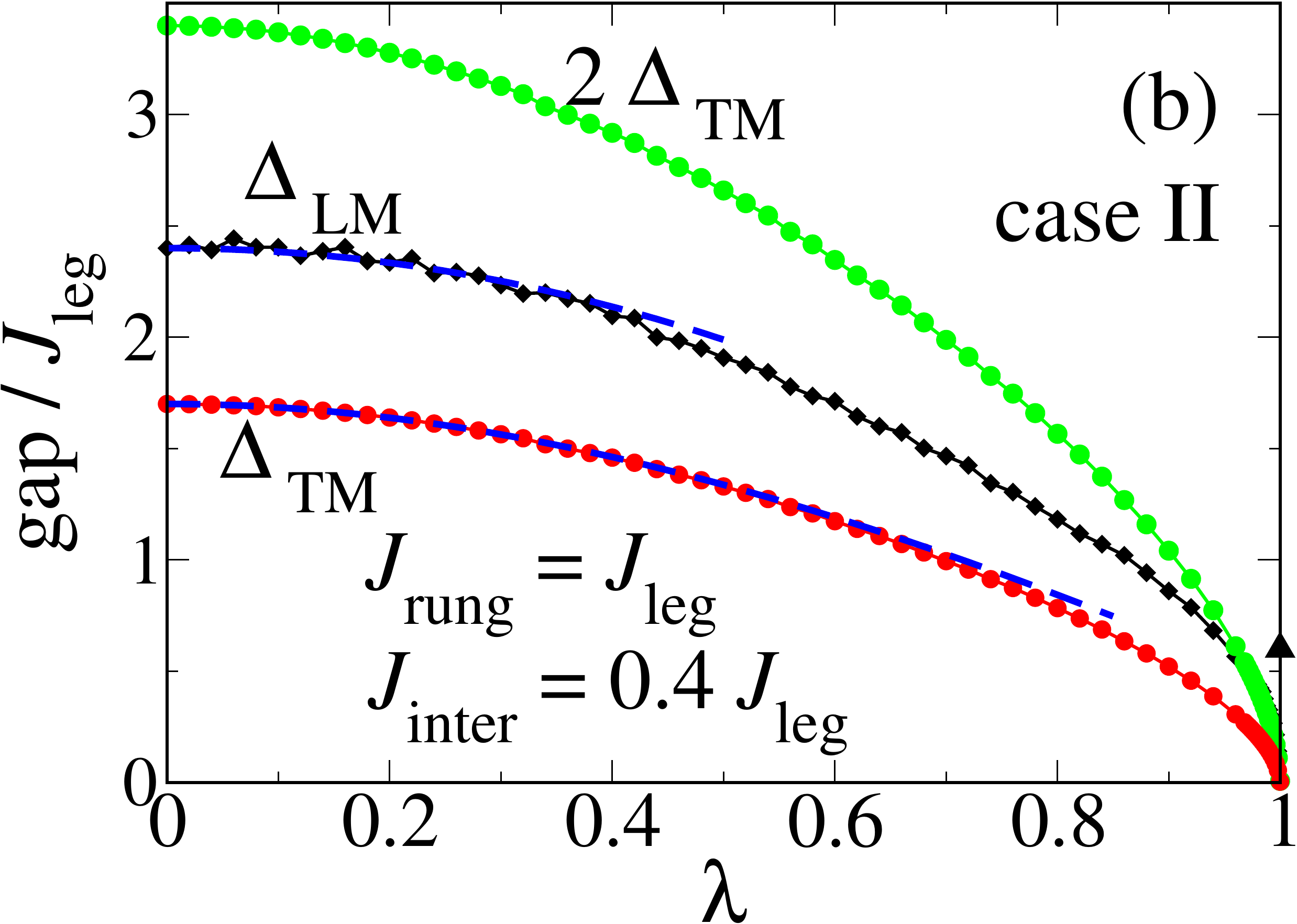}
\caption{ Excitation gaps  $\Delta_\mathrm{TM}$, $2 \Delta_\mathrm{TM}$ and $\Delta_\mathrm{LM}$ as functions of $\lambda$ at  $J_\mathrm{rung}=J_\mathrm{leg}$ for  (a)
 $J_\mathrm{inter} =0.228 J_\mathrm{leg}$ (case I), and  (b)  $J_\mathrm{inter} =0.4 J_\mathrm{leg}$ (case II). Dashed lines:  results 
 from series expansions.  Triangle in (b): position of the Higgs mode for $\lambda=1$ from the dynamic singlet structure factor. 
 }
\label{fig:gaps}
\end{figure}
 
From QMC data such as in Fig.~\ref{fig:skw}, we extract the full $\lambda$-dependence of both gaps in the thermodynamic limit~\cite{SM}, cf.~Fig.~\ref{fig:gaps}. Also shown in this figure are series expansion results~\cite{Hong17,SM,Dusuel10} up to order $\lambda^2$ ($\lambda^8$) for $\Delta_\mathrm{LM}$ ($\Delta_\mathrm{TM}$), which closely follow the QMC data up to intermediate values of $\lambda$.
For  case I, at $J_\mathrm{inter}/J_\mathrm{leg}=0.228$ [Fig.~\ref{fig:gaps}(a)], we identify the quantum critical point at $\lambda_c=0.964(2)$, where $\Delta_\mathrm{LM}$  closes. $\Delta_\mathrm{TM}$ stays finite across the transition, exhibiting an inflection point. While in the antiferromagnetic regime, $\lambda < \lambda_c$, the LM mode connects to a two-spin-flip bound state of the Ising limit, it forms the $S^z=0$ sector of the gapped triplon mode in the quantum disordered regime, 
which is degenerate with the TM mode of the transverse branch in the Heisenberg limit. The LM mode resides below the two-magnon continuum of energies above $2\Delta_\mathrm{TM}$ for all $\lambda$.
 For case II, at $J_\mathrm{inter}/J_\mathrm{leg}=0.4$  [Fig.~\ref{fig:gaps}(b)], the 
antiferromagnetic regime extends up to the Heisenberg limit, in which the TM gap closes. The softening of $\Delta_\mathrm{TM}$ effects the LM mode to merge into the two-magnon continuum, which we locate  to occur at $\lambda_m=0.96(2)$. Beyond this point, the detection of the Higgs mode is masked by the two-magnon continuum. Close to quantum criticality and in the Heisenberg limit ($\lambda=1$), one may nevertheless detect the Higgs mode through the scalar susceptibility in terms of the rung-based dynamic singlet structure factor~\cite{Lohoefer15, Lohoefer17,Qin17}. The position of the Higgs mode from this scalar response function is also shown in Fig.~\ref{fig:gaps}(b); it compares well to the energy of the LM mode near $\lambda_m$.

\begin{figure}[t]
\includegraphics[width=0.9\columnwidth]{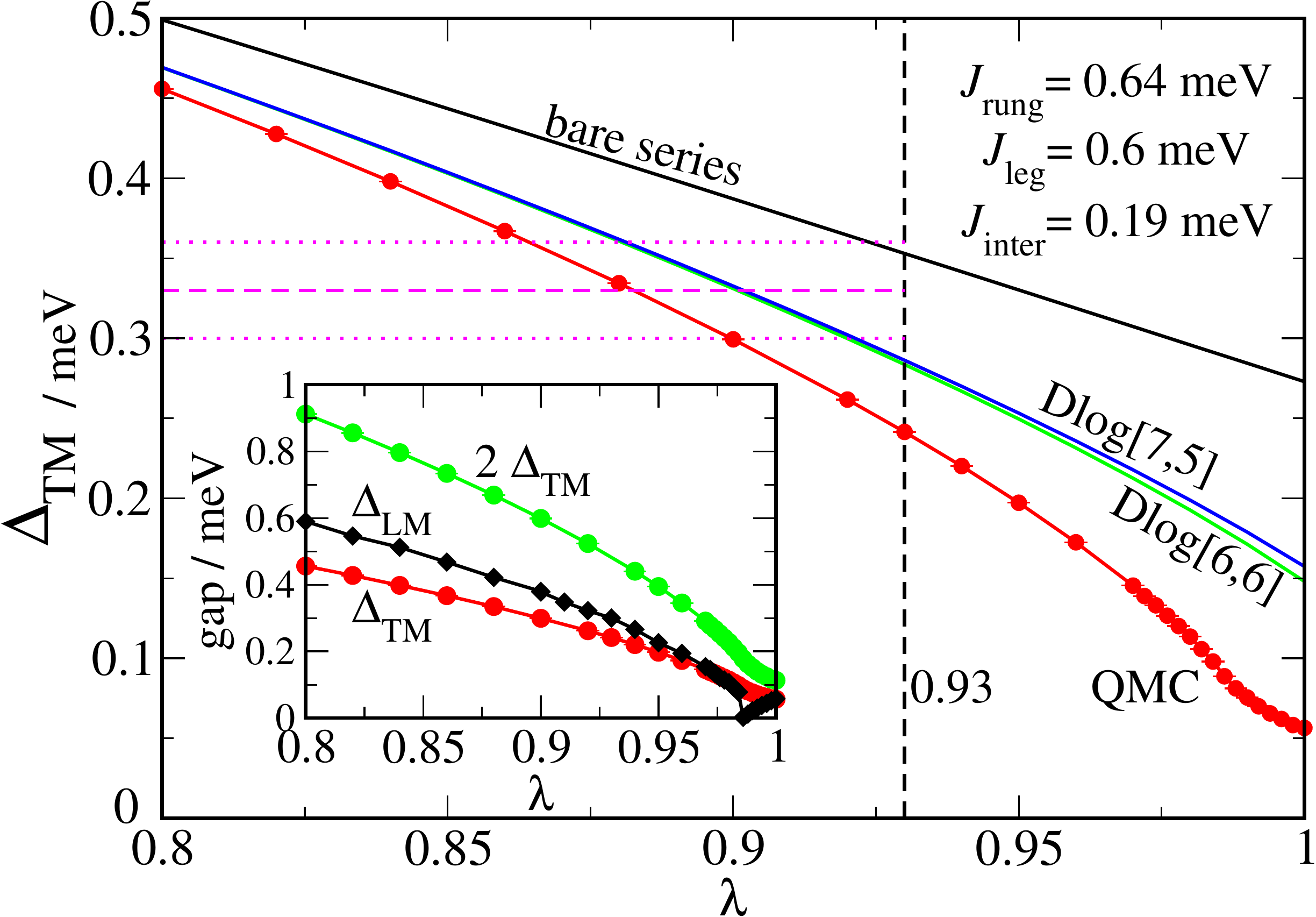}
\caption{Excitation gap $\Delta_\mathrm{TM}$ from QMC and pCUTS (bare order $\lambda^8$ series and Dlog-pad\'{e} approximants) 
as functions of $\lambda$ for 
$J_\mathrm{leg}=0.6$ meV,
 $J_\mathrm{rung}=0.64$ meV, and
 $J_\mathrm{inter} =0.19$ meV.
 Horizontal lines set the margin of the INS  estimate  $\Delta_\mathrm{TM}=0.33(3)$ meV for DLCB from Ref.~\cite{Hong17}. 
The inset shows $\Delta_\mathrm{TM}$, $\Delta_\mathrm{LM}$, and $2 \Delta_\mathrm{TM}$ 
as functions of  $\lambda$ as obtained from QMC.}
\label{fig:gapsexp}
\end{figure}

We next return to the theoretical modeling of the INS spectra for DLCB.
Since this compound  resides within the antiferromagnetically ordered regime of coupled spin-ladders, we first assess, to which of the two  cases (I or II) it belongs, according  to the effective description by the model in Eq.~(\ref{eq:model}). For this purpose, we performed QMC simulations for the set of previously estimated exchange couplings, but vary the anisotropy $\lambda$. We observe from Fig.~\ref{fig:gapsexp} that based on this parameter set, DLCB actually belongs to case I, i.e., for the estimated exchange couplings, $H$  resides within the quantum disordered regime at $\lambda=1$: The easy-axis anisotropy not only effects finite magnetic excitation gaps, it also leads  out of the quantum disordered regime. The presence of a quantum phase transition at $\lambda_c=0.989(1)$ (detected also by the antiferromagnetic structure factor~\cite{SM})  for this set of couplings was not noted in Ref.~\cite{Hong14,Hong17}, wherein  pCUT-based estimates of  $\Delta_\mathrm{TM}$ were used instead. As shown in Fig.~\ref{fig:gapsexp}, this approach does not reproduce  the inflection point in  
$\Delta_\mathrm{TM}$ at $\lambda_c$ and 
overestimates the gap in the relevant  parameter regime.
Therefore, the gap $\Delta_\mathrm{TM} \approx 0.24$ meV extracted from the QMC calculations at the previously estimated value of $\lambda= 0.93$ falls below the experimental margin for DLCB, i.e., a lower value of $\lambda$  is required to match  the experimental values of the gaps for the considered exchange coupling strengths.   
Agreement with the experimental estimates of the gaps within their error margins can be reached using a simple rescaling procedure: In order to satisfy Eq.~(\ref{eq:constraint}), a decrease in $\lambda$ 
requires a corresponding increase of the exchange coupling strengths. 
Here, we constrain to a uniform rescaling of all exchange constants for simplicity.
Using an interpolation of  the QMC data in Fig.~\ref{fig:gapsexp}~\cite{SM}, we obtain  
$J_\mathrm{leg}=0.619$ meV,
$J_\mathrm{rung}=0.660$ meV,
$J_\mathrm{leg}=0.196$ meV, and $\lambda=0.871$, 
for which $\Delta_\mathrm{TM}=0.360$ meV and $\Delta_\mathrm{LM}=0.457$ meV,  i.e.,  both values are
within the margins of the experimental estimates. 
We thus spared a fit of all four parameters of $H$ to the INS data, which is  rather expensive based on QMC calculations of the DSF.

 \begin{figure}[t]
\includegraphics[width=\columnwidth]{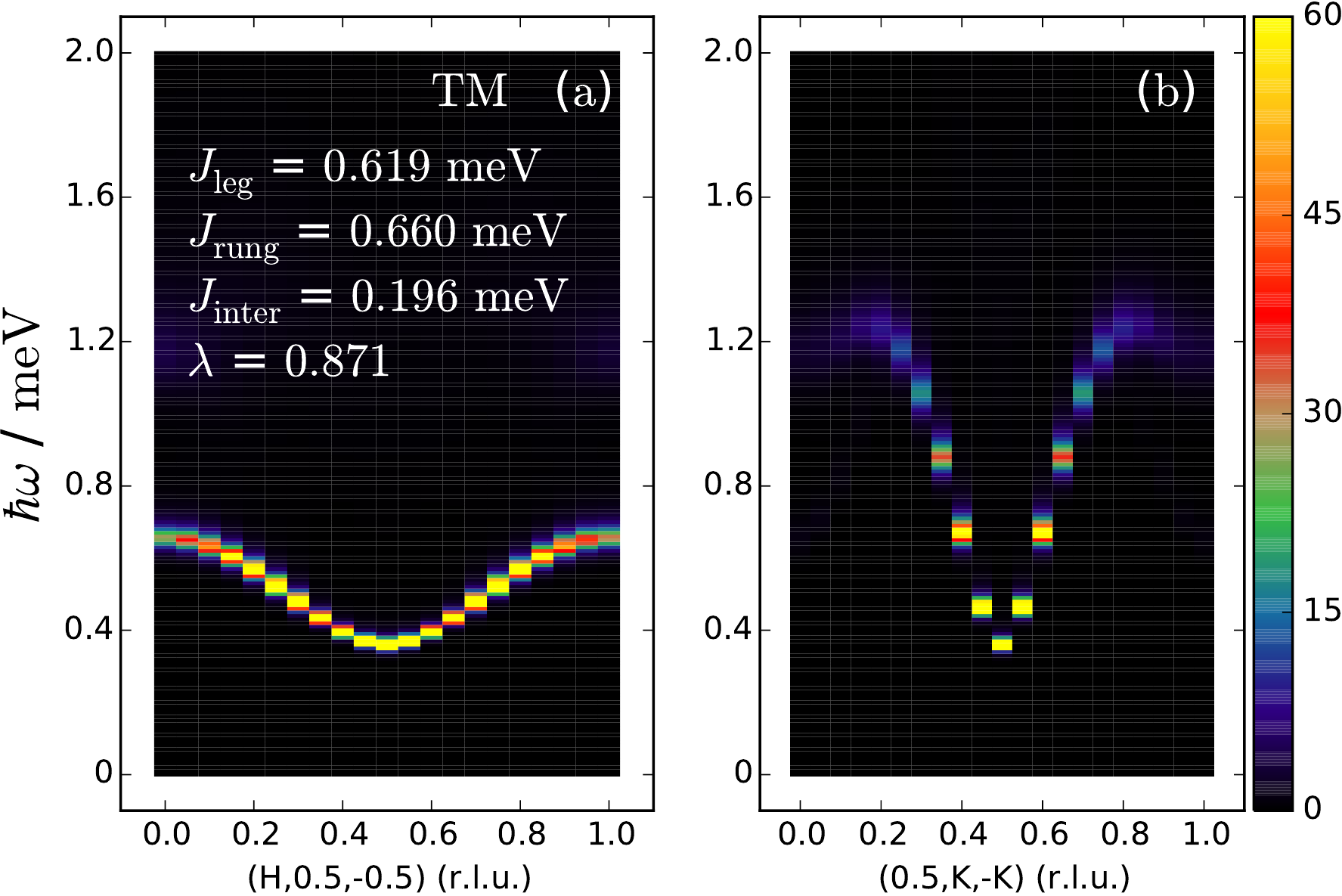}\vspace{3mm}
\includegraphics[width=\columnwidth]{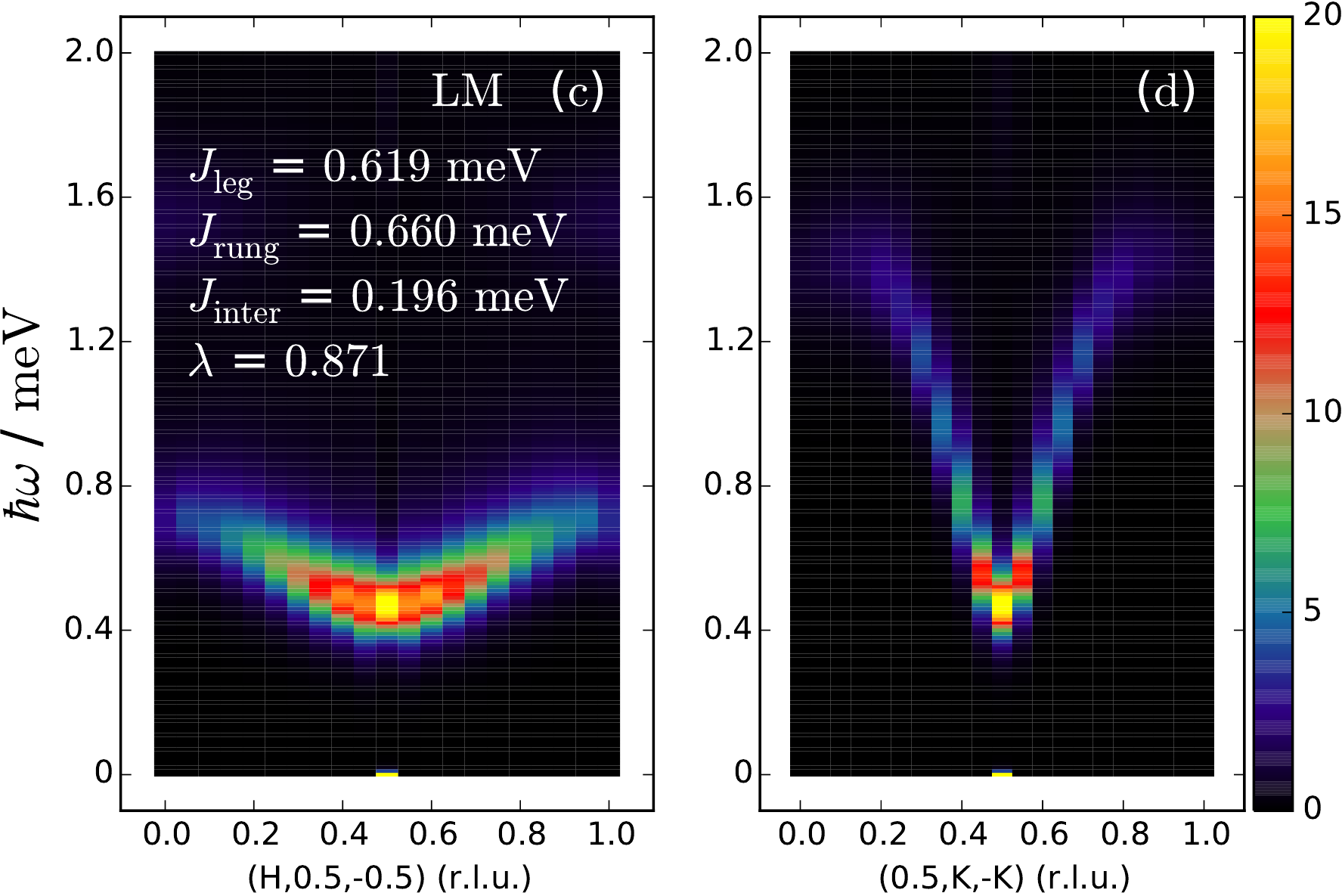}
\caption{Scattering spectra for DLCB as a function of energy and wave-vector transfer in the transverse (a, b) and longitudional (c, d) configurations, exhibiting the TM and LM modes, respectively. Data based on 
QMC simulations ($L=20$) of the 2D model $H$ for the displayed  parameters.}
\label{fig:expspec}
\end{figure}

Based on this consistent identification of a single set of model parameters for DLCB, we finally performed QMC simulations to calculate the corresponding DSF. To allow for a  direct comparison to the INS results presented in Ref.~\cite{Hong17}, we transformed the QMC spectra~\cite{SM} to the crystal and scattering geometry for DLCB~\cite{Hong17}. The resulting  scattering spectra along the specific wave-vector transfers considered in  Ref.~\cite{Hong17} are shown in Fig.~\ref{fig:expspec} for both  polarization directions. They correspond  to the polarized INS data shown in Fig.~4 of Ref.~\cite{Hong17}. In addition to the excitation gaps and the overall distribution of the spectral weight, the calculated spectra 
also account for the  bandwidth  observed in the INS spectra in the LM scattering channel at the zone boundary,  which was overestimated in the harmonic BOT  approach from Ref.~\cite{Hong17}.

Hence, we  demonstrated the feasibility, using state-of-the-art QMC simulation techniques, to formulate a quantitative theory for the  spin dynamics of near-quantum-critical 2D quantum magnets, directly
exposing
 the two-magnon bound-state  nature of the stable Higgs mode excitation observed in recent INS experiments on DLCB. In the easy-axis regime, this excitation is stabilized due to  the upwards-shifted support of the two-magnon continuum, well above the Higgs mode's excitation gap. The Higgs mode merges into this continuum only very close to the Heisenberg limit within the antiferromagnetic regime, beyond the quantum critical point. We anticipate our unbiased QMC approach to provide a 
quantitative understanding to the  quantum spin dynamics also in other near-quantum-critical 2D magnetic compounds. 

\acknowledgments

{\it Acknowledgments.} We thank Tao Hong for valuable discussions. This work was supported by the Deutsche 
Forschungsgemeinschaft (DFG) under Grants No. FOR 1807 and No. RTG 1995, as well as  the National Natural Science Foundation of China  under Grant No. 11504067.
We thank the IT Center at RWTH Aachen University and 
the J\"ulich Supercomputing Centre for access to computing time through JARA-HPC.

%
%
\end{document}


\title{Supplemental Material for 
``The Higgs Mode of   Planar Coupled Spin-Ladders
and  its Observation in  C$_9$H$_{18}$N$_2$CuBr$_4$''}

\author{T. Ying, K. P. Schmidt, S. Wessel}

\maketitle

\noindent
{\bf S1. Details on  the DSF formula} 
\medskip

To obtain  the DSF from QMC, it is convenient to express the
DSF in a form that is explicit in the  unit cell decomposition. 
With reference to the inset of Fig.~1 in the main text,  we denote
the vector that connects two unit cells along the ladder (leg) direction  by 
$\bm{a}_1$, and correspondingly define the vector $\bm{a}_2$ along the rung direction.
As  mentioned in the main text,  we fix units such that the unit cells length are equal to
1 in both directions.
The vector 
$\bm{\delta}$ points from the lower to the upper spin inside a unit cell, cf. the inset of Fig.~1 in the main text, and equals $\bm{\delta}=(0, 1/2)$ for this geometry. 
In the following, $\mu$ and $\nu$ label unit cells and $\bm{S}_{\mu1}$ ($\bm{S}_{\mu2}$) the lower (upper) spin in the
$\mu$-th unit cell, respectively. Furthermore, we fix the positions of the spins such
that $\bm{r}_{\mu1}=\bm{R}_\mu$ and $\bm{r}_{\mu2}=\bm{R}_\mu+\bm{\delta}$. Here,  $\bm{R}_\mu$  denotes the position vector of the $\mu$-th unit cell. The number of spins $N$ and the number of unit cells $N_u$ are
related by $N=2 N_u$.
In terms of $\bm{\delta}$, we
then obtain
\begin{equation}\label{eqS:s1}
S_\mathrm{L}(\omega,\bm k) = \cos^2(\bm{k}\cdot\bm{\delta}/2)S_\mathrm{L}^s(\omega,\bm k) +  \sin^2(\bm{k}\cdot\bm{\delta}/2)S_\mathrm{L}^a(\omega,\bm k),
\end{equation}
and 
\begin{equation}\label{eqS:s2}
S_\mathrm{T}(\omega,\bm k) = \cos^2(\bm{k}\cdot\bm{\delta}/2)S_\mathrm{T}^s(\omega,\bm k) +  \sin^2(\bm{k}\cdot\bm{\delta}/2)S_\mathrm{T}^a(\omega,\bm k),
\end{equation}
from the symmetric and antisymmetric structure factors
\begin{eqnarray}
\label{eq:SfactorU}
S_\mathrm{L}^{s/a}(\omega,\bm k) &=& \frac{1} {2 N_u}\int{dt}{\sum_{\mu,\nu}}e^{{\rm i}\left(\hbar\omega t-\bm{k} \cdot (\bm{R}_\mu-\bm{R}_\nu)\right)}\\
&& \times\langle ({S}^z_{\mu1}(t) \pm {S}^z_{\mu2}(t) )  \cdot ({S}^z_{\nu1}(0)\pm {S}^z_{\nu2}(0)  ) \rangle ,\nonumber
\end{eqnarray}
and
\begin{eqnarray}
\label{eq:SfactorU}
S_\mathrm{T}^{s/a}(\omega,\bm k) &=& \frac{1} {2 N_u}\int{dt}{\sum_{\mu,\nu}}e^{{\rm i}\left(\hbar\omega t-\bm{k} \cdot (\bm{R}_\mu-\bm{R}_\nu)\right)}\\
&&\times(\: \langle ({S}^+_{\mu1}(t) \pm {S}^+_{\mu2}(t) )  \cdot ({S}^-_{\nu1}(0)\pm {S}^-_{\nu2}(0)  ) \rangle \nonumber\\
&&   \:\:\:+\langle ({S}^-_{\mu1}(t) \pm {S}^-_{\mu2}(t) )  \cdot ({S}^+_{\nu1}(0)\pm {S}^+_{\nu2}(0)  ) \rangle \:)  ,\nonumber
\end{eqnarray}
with respect to the interchange of the two legs of each ladder. 
These are conveniently obtained separately from QMC simulations.

\bigskip\smallskip
\noindent
{\bf S2. Finite-size scaling analysis}
\medskip

\begin{figure}[t]
\includegraphics[width=0.9\columnwidth]{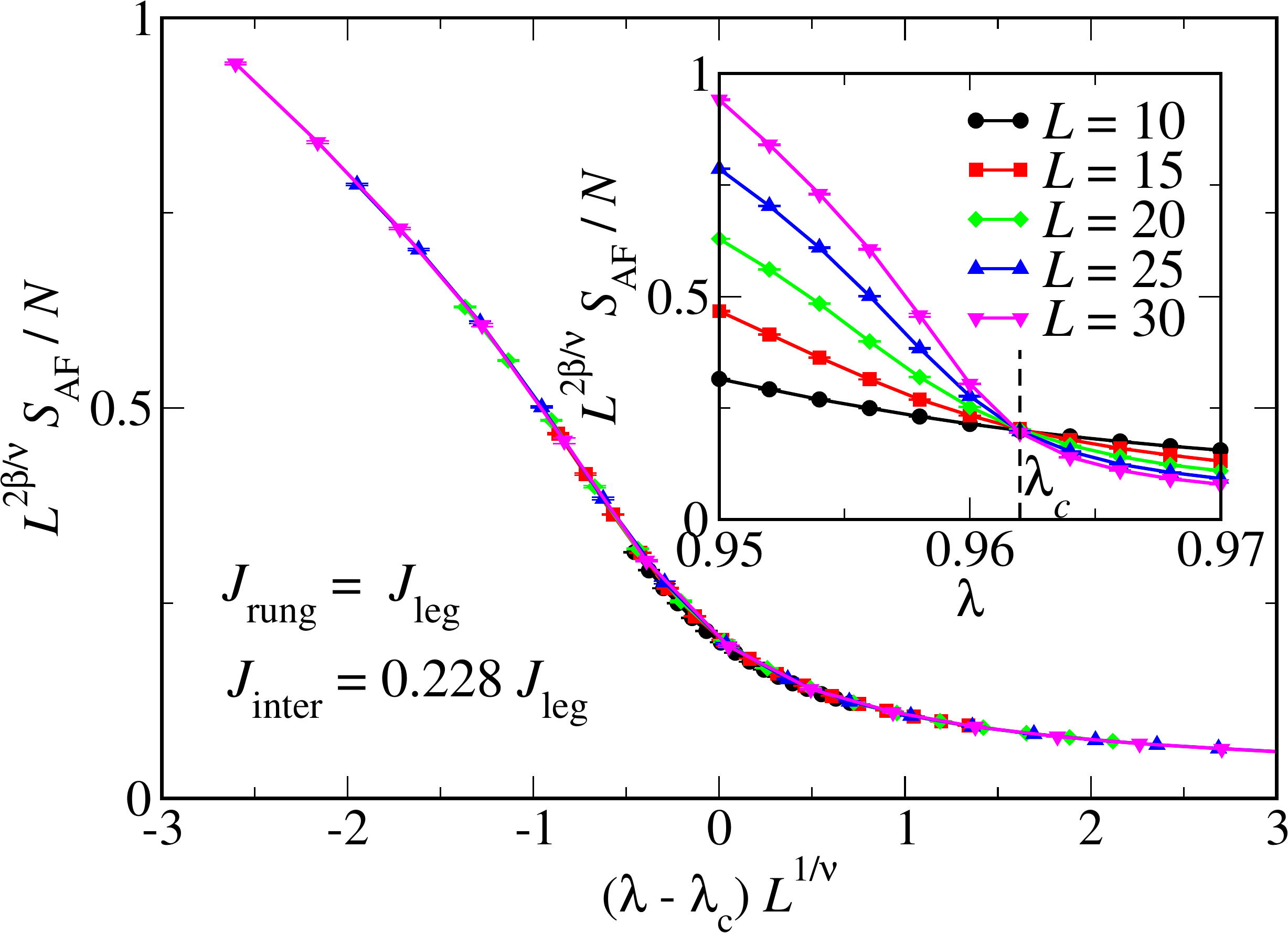}
\caption{Finite-size scaling  of  $S_\mathrm{AF}$ at the quantum phase transition  for a 2D array of coupled ladders with $J_\mathrm{rung}=J_\mathrm{leg}$, and  $J_\mathrm{inter}=0.228J_\mathrm{leg}$. The critical value of $\lambda_c=0.964(2)$ is extracted as the crossing point of the appropriately rescaled finite-size values of  $S_\mathrm{AF}$ (inset). The main panel shows the corresponding data collapse of the finite-size data. The critical exponents $\beta=0.326419(3)$ and $\nu=0.629971(4)$ from conformal bootstrap calculations~\cite{Elshowk} for the 3D Ising universality class were employed.}
\label{figS:scaling1}
\end{figure}

\begin{figure}[t]
\includegraphics[width=0.9\columnwidth]{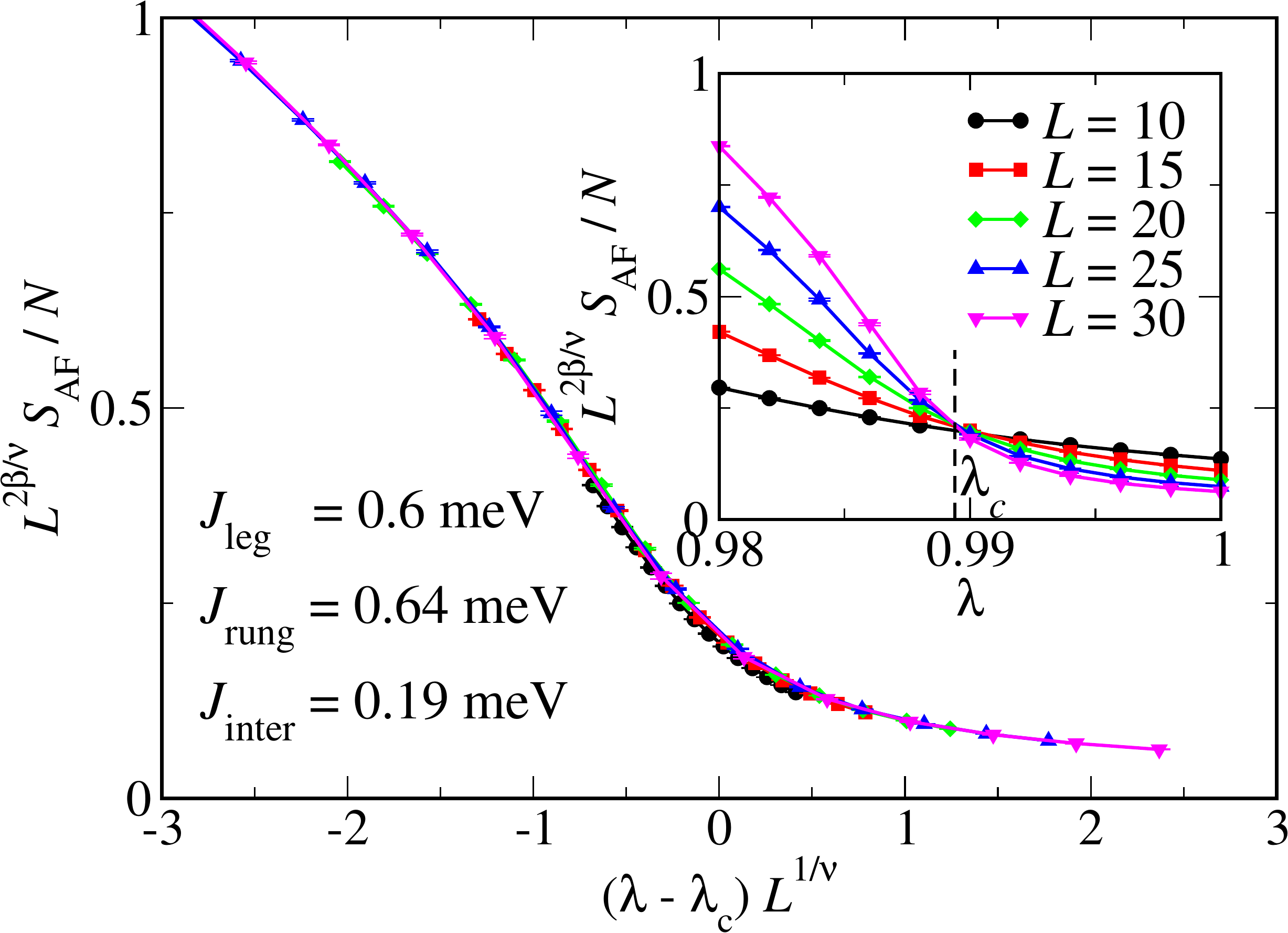}
\caption{Finite-size scaling  of  $S_\mathrm{AF}$ at the quantum phase transition  for a 2D array of coupled ladders with $J_\mathrm{rung}=0.64$ meV, $J_\mathrm{leg}=0.6$ meV, and  $J_\mathrm{inter}=0.19$ meV. The critical value of $\lambda_c=0.989(1)$ is extracted as the crossing point of the appropriately rescaled finite-size values of  $S_\mathrm{AF}$ (inset). The main panel shows the corresponding data collapse of the finite-size data. The critical exponents $\beta=0.326419(3)$ and $\nu=0.629971(4)$ from conformal bootstrap calculations~\cite{Elshowk} for the 3D Ising universality class were employed.}
\label{figS:scaling2}
\end{figure}

We performed a standard finite-size scaling analysis of the antiferromagnetic structure factor, which is given as $S_\mathrm{AF}=\langle ({S}^z_{\bm{k}_\mathrm{AF}})^2 \rangle$ in our system.
Here, $\bm{k}_\mathrm{AF}=(\pi,2\pi)$ is the antiferromagnetic ordering vector.
In accord with a dynamical critical exponent $z=1$, we scaled the temperature as
$T/J_\mathrm{leg}=4L$ with the system size for these QMC simulations. 
This analysis is shown in Fig.~\ref{figS:scaling1} for the case of $J_\mathrm{rung}=J_\mathrm{leg}$, $J_\mathrm{inter}/J_\mathrm{leg}=0.228$, 
and in  Fig.~\ref{figS:scaling2} for the case of 
$J_\mathrm{rung}=0.64$ meV, $J_\mathrm{leg}=0.6$ meV, and  $J_\mathrm{inter}=0.19$ meV, considered in the main text. 

 \bigskip\smallskip
\noindent
{\bf S3. Determination of the excitation gaps}
\medskip

\begin{figure}[t]
\includegraphics[width=0.9\columnwidth]{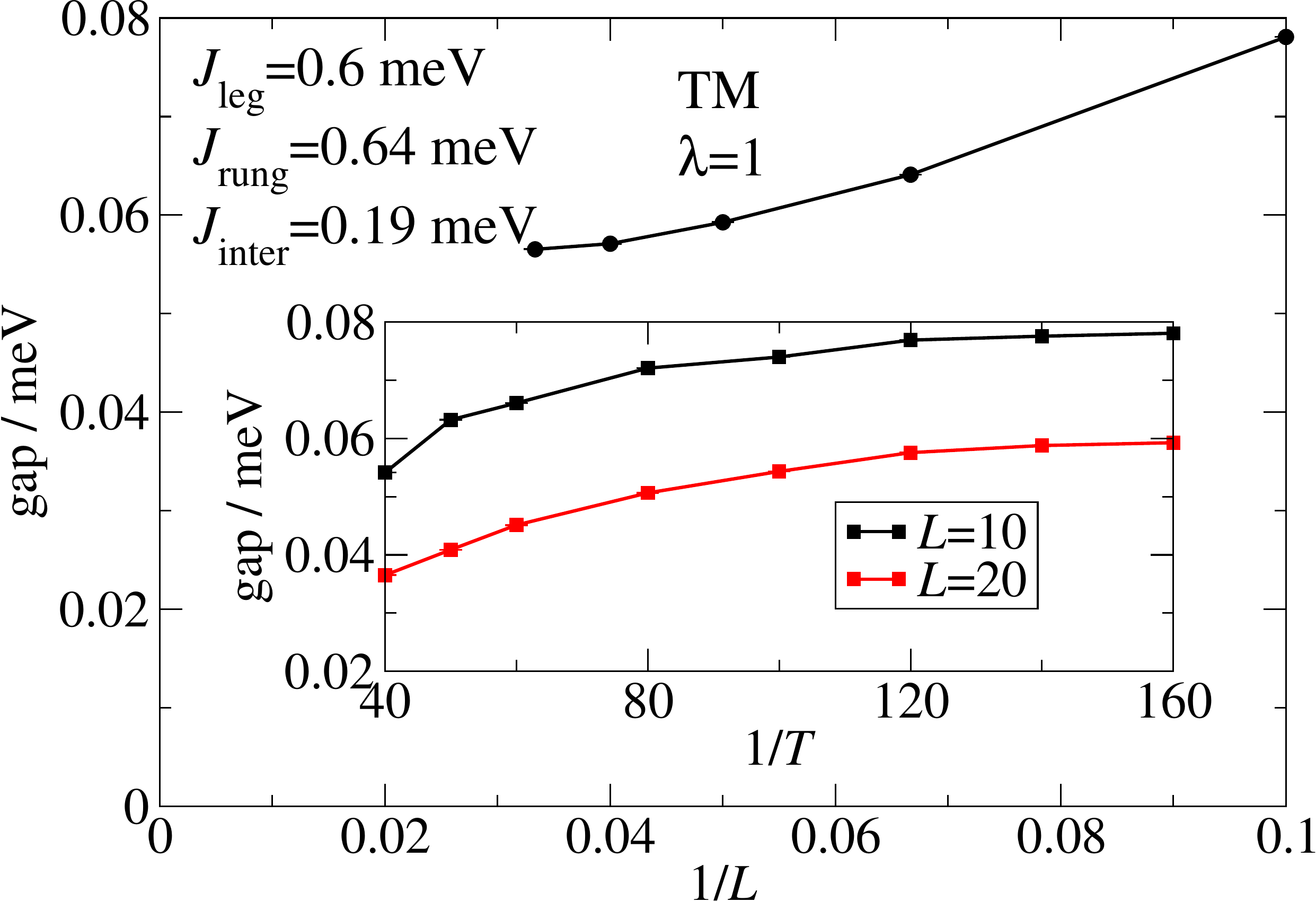}
\caption{System size and temperature dependence (inset) of the excitation gap 
$\Delta_\mathrm{TM}$ for the played set of model parameters. This procedure was used to obtain the values given in Figs. 2 and 3 of the main text.
}
\label{figS:gapesti}
\end{figure}

We used the following standard procedure to estimate the gaps given in Figs.~3 and 4 of the main text:
In order to determine the excitation gaps for a given finite lattice, we first extracted 
$\Delta_\mathrm{TM}$ and
$\Delta_\mathrm{LM}$ as the energy of the peak  in the corresponding DSF at  momentum $\bm{k}_\mathrm{AF}$, based on for the DSF such as shown in Fig.~2 of the main text. Alternatively, we also obtained  estimates for 
 $\Delta_\mathrm{TM}$  from the corresponding imaginary-time displayed transverse correlation function at momentum $\bm{k}_\mathrm{AF}$, which provided consistent values. 
For a given set of parameters of $H$, we then examined the temperature dependence of the gap estimates to ensure that we obtain their ground state values. Finally, we examined the system size dependence of the lowest-temperature gaps  
to ensure that the quoted values are also converged in terms of system size. This procedure is shown explicitly in Fig.~\ref{figS:gapesti}. Based on such an analysis, we finally obtained the values given in Figs.~3 and 4 of the main text.

\bigskip\smallskip
\noindent
{\bf S4. Series expansion calculations} 
\medskip

In this subsection we present the necessary information on the series expansion performed about the Ising limit. We note that this expansion is different to the one used in Ref.~\cite{Hong14}, where part of the classical Ising Hamiltonian, corresponding to the density-density interaction in terms of hardcore bosons, has been included in the perturbation so that the remaining unperturbed part has an equidistant spectrum. 
Here, the unperturbed Hamiltonian is the full Ising interaction
\begin{equation}
 \label{Eq::H0_Ising}
{H}_0 = \frac{1}{4}\sum_{\alpha\in\{{\rm rung}, {\rm leg},{\rm int}\}} J_\alpha\sum_{\alpha-{\rm bonds}} \sigma^z_i \sigma^z_j\quad,
\end{equation}
with antiferromagnetic interactions $J_\alpha$. The two  ground states of $H_0$ are the two N\'{e}el-ordered states. For the perturbative expansion inside this ordered phase we choose one of them and perform a sublattice rotation about the $x$-axis, with $\sigma^x\rightarrow\sigma^x$, $\sigma^y\rightarrow -\sigma^y$, and $\sigma^z\rightarrow\sigma^z$. As a consequence, Eq.~\eqref{Eq::H0_Ising} is transformed to
\begin{equation}
 \label{Eq::H0_Ising_mapped}
 \tilde{{H}}_0 = -\frac{1}{4}\sum_{\alpha\in\{{\rm rung}, {\rm leg},{\rm int}\}} J_\alpha\sum_{\alpha-{\rm bonds}} \sigma^z_i \sigma^z_j
\end{equation}
and the ordered ground states are now the two ferromagnetic ones with a ground-state energy
\begin{equation}
  E_0 = -\frac{N}{8}\left( J_{\rm rung}+2J_{\rm leg} + J_{\rm int}\right)\quad .
\end{equation}
The elementary excitation is a single spin-flip relative to the ferromagnetic ground state. The corresponding one-particle energy is $E_0$ plus the energy costs of the four antiferromagnetic configurations
\begin{equation}
  E_1 = E_0+\frac{1}{2}\left( J_{\rm rung}+2J_{\rm leg} + J_{\rm int}\right)\quad .
\end{equation}
In the following, we concentrate on so-called two-particle bound states, which represent configurations with two nearest-neighbor spin flips. There are three different types of such bound states, depending on whether the double flip is located on a $J_{\rm rung}$-, $J_{\rm leg}$-, or $J_{\rm int}$-bond. The unperturbed energies are given by
\begin{eqnarray}
  E_2^{\rm rung} &=& E_0+\frac{1}{2}\left( 4J_{\rm rung} + 2J_{\rm int}\right)\\
  E_2^{\rm leg} &=& E_0+\frac{1}{2}\left( 2J_{\rm rung} + 2J_{\rm leg} + 2J_{\rm int}\right)\\
  E_2^{\rm int} &=& E_0+\frac{1}{2}\left( 2J_{\rm rung} + 4J_{\rm leg} \right)\quad .
\end{eqnarray}
The perturbation is the transverse part of the Hamiltonian $H$, given by 
\begin{equation}
 \label{Eq::V}
 {V} = \frac{\lambda}{4}\sum_{\alpha\in\{{\rm rung},{\rm leg},{\rm int}\}} J_\alpha\sum_{\alpha-{\rm bonds}} \left( \sigma^x_i \sigma^x_j+ \sigma^y_i \sigma^y_j\right) \quad.
\end{equation}
Applying the sublattice rotation to this perturbation yields
 \begin{equation}
 \label{Eq::V_mapped}
 \tilde{{V}} = \frac{\lambda}{2}\sum_{\alpha\in\{{\rm rung},{\rm leg},{\rm int}\}} J_\alpha\sum_{\alpha-{\rm bonds}} \left( \sigma^+_i \sigma^+_j+ \sigma^-_i \sigma^-_j\right) \quad.
\end{equation}
Locally, on a given nearest-neighbor bond, this perturbation  therefore gives  a finite matrix element only if it acts on a ferromagnetic configuration such that both involved spins can be flipped. Such a double-flip in any case changes the energy of the six bonds attached to that bond. We focus on the case $J_{\rm rung}=J_{\rm leg}= J$ with $J> J_{\rm int}$, i.e.~one has $E_2^{{\rm rung}}=E_2^{{\rm leg}}=E_2$ and the two-particle bound state on an $J_{\rm int}$-bond has a higher energy. Next, we perform second-order perturbation theory on the degenerate two-particle bound-state manifold with energy $E_2$. Introducing the projector on this subspace $\mathcal{P}$ and $\mathcal{Q}\equiv 1-\mathcal{P}$, the perturbative expansion reads on the operator level 
 \begin{equation}
 \label{Eq::taka}
 \tilde{{H}}_0+\mathcal{P}\tilde{{V}}\mathcal{P}+ \mathcal{P}\tilde{{V}}\mathcal{Q}\,\frac{1}{E_2-\tilde{{H}}_0} \mathcal{Q}\,\tilde{{V}}\mathcal{P} \quad.
\end{equation}
We first note that the first-order contribution vanishes exactly, since the perturbation $\tilde{\mathcal{V}}$ changes the spin-flip number always by two, and it is therefore the second-order contribution which gives finite matrix elements between different bound state configurations. The corresponding states are denoted by $|\mathbf{r}_{\nu},\alpha\rangle$, where $\mathbf{r}_{\nu}$ denotes the position of the $\nu$-th unit cell containing two spins forming a $J_{\rm rung}$-link and $\alpha\in\{{\rm rung},{\rm up}, {\rm down}\}$ represent the three different types of bound states attached to the unit cell ${\nu}$. Indeed, $\alpha={\rm rung}$ is the bound state on that $J_{\rm rung}$-link and $\alpha={\rm up}$ ($\alpha={\rm down}$) is the bound state on the upper (lower) $J_{\rm leg}$-bond attached to ${\nu}$. These three bound states per unit cell  ${\nu}$ can be combined to the three-dimensional vector 
\begin{equation}
\vec{v}_{{\nu}}=\big( |\mathbf{r}_{\nu},{\rm rung}\rangle , |\mathbf{r}_{\nu},{\rm up}\rangle, |\mathbf{r}_{\nu},{\rm down}\rangle \big)\quad .
\end{equation}
The second-order perturbation corresponds to hopping processes of these bound states, either with or without change of bound-state type $\alpha$. Exploting the translational symmetry by applying the Fourier transform
\begin{equation}
\vec{v}_{\mathbf{k}}=\frac{1}{2N}\sum_{{\nu}} \exp({\rm i}\mathbf{k}\cdot\mathbf{r}_{\nu})\, \vec{v}_{{\nu}}\quad ,
\end{equation}
we can write the effective Hamiltonian in the bound-state subspace up to second-order perturbation theory as
\begin{equation}
 \vec{v}_{\mathbf{k}}^\dagger\; \Omega(\mathbf{k})\;\vec{v}_{\mathbf{k}}\label{Eq::eff_ham}
 \quad,
\end{equation}
where $\Omega(\mathbf{k})$ is a $3 \times 3$ matrix containing the zeroth-order bound-state energies and the Fourier transform of all hopping amplitudes. Diagonalizing this matrix yields the three bound-state dispersions. We find that the gap $\Delta_{\rm LM}$ of the low-energy band has a momentum $\mathbf{k}=(\pi,\pi)$. 
In units of  $J=1$, we  then obtain, up to $O(\lambda^2)$,
%
%
%
\begin{eqnarray}
  \Delta_{\rm LM}= E_2 - \frac{\lambda^2}{4} \Big(
   \frac{2}{3}
   &&+ \frac{1}{2-2 J_{\rm int}}
   + \frac{20}{1+2J_{\rm int} }  
   + \frac{4}{4-J_{\rm int}}\nonumber\\
   && - \frac{18}{2+J_{\rm int}} 
   + \frac{1}{2-2J_{\rm int}}
\Big).
\end{eqnarray}
As shown in Fig. 3 of the main text, this results compares well to the QMC results up to intermediate values of $\lambda$. 

\bigskip\smallskip
\noindent
{\bf S5. Estimation of the  parameters for DLCB} 
\medskip

In order to estimate the coupling parameters relevant for DLCB based on the Hamiltonian $H$, 
we used the fact that in the regime of $0.84<\lambda<0.9$ in Fig.~4 of the main text, one can well fit the QMC data for $\Delta_\mathrm{TM}$ and  $\Delta_\mathrm{LM}$ to  approximative linear forms, 
$\Delta_\mathrm{TM}(\lambda)=1.822 - 1.6913 \lambda$, and 
$\Delta_\mathrm{LM}(\lambda)=2.3751 - 2.2182 \lambda$ (using meV units), respectively.
For a value of $\lambda$ below 0.93, one needs to increase the exchange coupling constants by a factor $f(\lambda)=1.93/(1+\lambda)$ due to Eq. (2) of the main text, and we require that $f(\lambda)(1.822 - 1.6913 \lambda) = 0.33(3)$ from the experimental value of $\Delta_\mathrm{TM}$. For a given value on the right-hand-site of the last equation, one can then readily solve this equation for $\lambda$. In particular, using 0.36, at the top of the error margin, we thereby  obtain the values given in the main text. 

\begin{figure}[t]
\includegraphics[width=0.9\columnwidth]{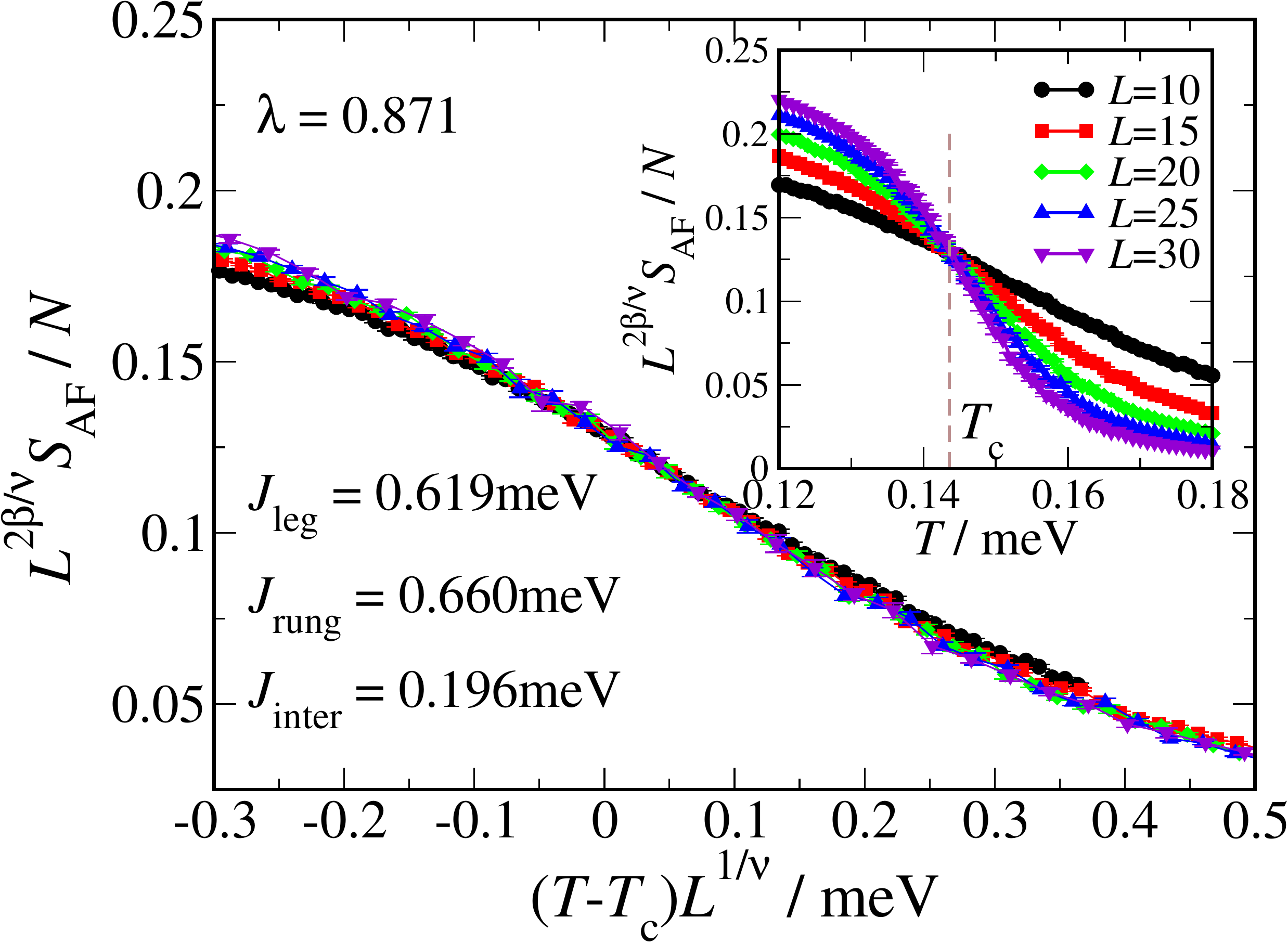}
\caption{Finite-size scaling  of  $S_\mathrm{AF}$ at the thermal phase transition for $J_\mathrm{rung}=0.660$ meV, $J_\mathrm{leg}=0.619$ meV,   $J_\mathrm{inter}=0.196$ meV, and $\lambda=0.871$, as estimated for the DLCB compound. The critical value of $T_c=0.1436$ meV is extracted as the crossing point of the appropriately rescaled finite-size values of  $S_\mathrm{AF}$ (inset). The main panel shows the corresponding data collapse of the finite-size data, which shows that corrections to the leading finite-size scaling affect the data on the accessed system sizes. The exact critical exponents $\beta=0.125$ and $\nu=1$ for the 2D Ising universality class were employed.}
\label{figS:scaling3}
\end{figure}

For this set of parameters, we also performed a standard finite-size scaling analysis in order to estimate the ordering temperature $T_c$, below with long-range antiferromagnetic order sets in, cf. Fig.~\ref{figS:scaling3}. From the crossing point analysis shown in the inset of Fig.~\ref{figS:scaling3}, we obtain an estimate of $T_c=0.1436$ meV, corresponding to $T_c=1.666$ K. This value compares well to the experimental value of $1.99(2)$ K for DLCB~\cite{Hong14}, given that weak interlayer couplings further enhance low-temperature ordering tendencies (these couplings are not included in the  Hamiltonian $H$ for a single layer).  
%
%

\bigskip\smallskip
\noindent
{\bf S6. Transforming to the  geometry for DLCB} 
\medskip

\begin{figure}[t]
\includegraphics[width=0.9\columnwidth]{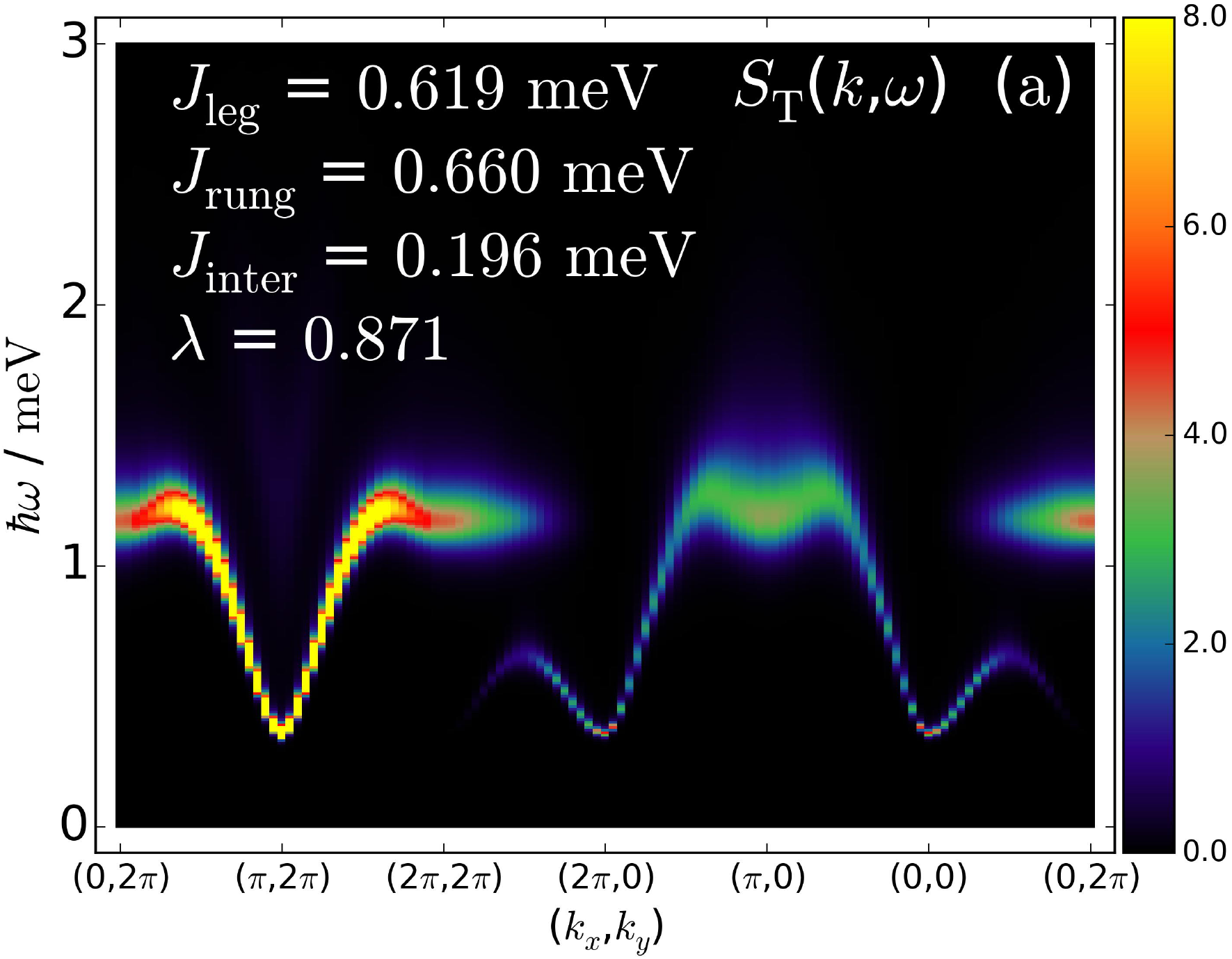}
\includegraphics[width=0.9\columnwidth]{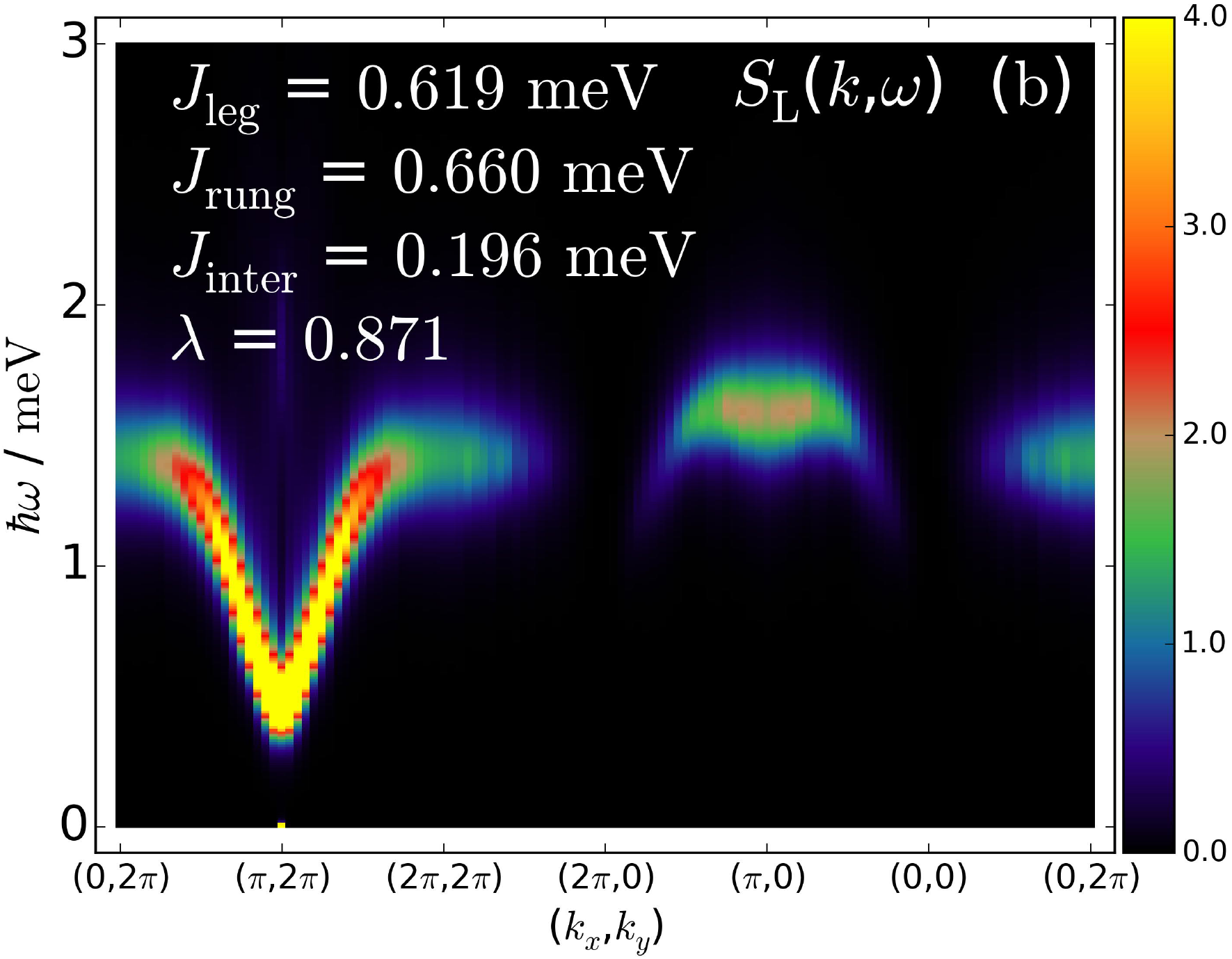}
\caption{DSF for DLCB,  (a) $S_\mathrm{T}(\bm{k},\omega)$,  and (b) $S_\mathrm{L}(\bm{k},\omega)$, along the indicated path in momentum space, obtained from QMC simulations with $L=20$ at $T=0.14$ K.}
\label{figS:skw}
\end{figure}

In Fig.~\ref{figS:skw}, we present the DSF of the model Hamiltonian $H$ for the above estimated coupling parameters for DLCB along the indicated path through momentum space. To probe the spin dynamics in the ground state, we considered a sufficiently low temperature $T=0.14$ K, well below the ordering temperature $T_c$ estimated in the previous section. 

In order to transform the QMC spectra to the experimental scattering geometry, we proceeded as follows, based on the results of Refs.~\cite{Hong14,Hong16,Hong17}:
We denote by $\bm{a}$, $\bm{b}$, and $\bm{c}$ the lattice vectors  of the compound DLCB, and by $\bm{a}^*$, $\bm{b}^*$, and $\bm{c}^*$ the corresponding reciprocal lattice vectors. 
The lattice vectors $\bm{a}_1$ and $\bm{a}_2$ to be used in our planar model for DLCB are given  explicitly as 
$\bm{a}_1=\bm{b}$, and $\bm{a}_2=\bm{a}-\bm{c}$.
Furthermore, we can also express $\bm{\delta}$ in terms of the lattice vectors as $\bm{\delta}=\delta_a \bm{a} + \delta_b \bm{b} + \delta_c \bm{c}$; the  coefficients   are found to  be 
$\delta_a=0.54832$,
$\delta_b=-0.14964$, and
$\delta_c=-0.56776$
from the lattice structure for DLCB.

We are interested in the scattering spectra 
with transfer wave-vector  $\bm{q}=H\bm{a}^*+K\bm{b}^*+L\bm{c}^*$,
for both the longitudinal (LM mode)  and the transverse (TM mode) configuration,
\begin{equation}
I_\mathrm{L/T}(\omega, \bm q)=I_\mathrm{L/T}(\omega, H, K , L)
\end{equation}
The zero-field alignment direction of the magnetic moments is parallel to $\bm{c}^*$. We define a corresponding unit vector
$\hat{\bm z}=\bm{c}^*/|\bm{c}^*|$. 
Then, $I_\mathrm{L/T}$ is computed in terms of the DSF from
\begin{eqnarray}\nonumber
I_\mathrm{L/T}(\omega,H,K,L)&=&\left( \!1- \frac{q^2_z}{\bm{q}^2} \!\right)
\left| \frac {g} {2} F\!\left(\frac{|\bm{q}|} {4\pi}\right) \!\right|^2 
\!\!S_\mathrm{L/T}(\omega, \bm{k}_{HKL}),
\end{eqnarray}
where $q_z=\hat{\bm z}\cdot \bm{q}$ is the $z$-component of $\bm{q}$, and $\bm{k}_{HKL}$ is the two-component vector
$\bm{k}_{HKL}=({2\pi K},{2\pi (H-L)})$.
Furthermore, 
 $F(\cdot)$ is the magnetic form factor of the Cu${}^{2+}$ ions ($g=2.12$), 
and finally, from Eqs.~(\ref{eqS:s1}) and (\ref{eqS:s2}), we obtain 
\begin{eqnarray}
S_\mathrm{L/T}(\omega, \bm{k}_{HKL})&=&\frac{1}{2}\left( S^{s}_\mathrm{L/T}(\omega, \bm{k}_{HKL}) + S^{a}_\mathrm{L/T}(\omega, \bm{k}_{HKL}) \right)\nonumber \\
& & + \frac{\cos \left( 2\pi(H\delta_a+K\delta_b+L\delta_c) \right)}{2}\nonumber \\
& &\: \times \left( S^{s}_\mathrm{L/T}(\omega, \bm{k}_{HKL}) - S^{a}_\mathrm{L/T}(\omega, \bm{k}_{HKL}) \right). \nonumber
\end{eqnarray}